# A Conversation with Shayle R. Searle

**Martin T. Wells**

*Abstract.* Born in New Zealand, Shayle Robert Searle earned a bachelor's degree (1949) and a master's degree (1950) from Victoria University, Wellington, New Zealand. After working for an actuary, Searle went to Cambridge University where he earned a Diploma in mathematical statistics in 1953. Searle won a Fulbright travel award to Cornell University, where he earned a doctorate in animal breeding, with a strong minor in statistics in 1959, studying under Professor Charles Henderson. In 1962, Cornell invited Searle to work in the university's computing center, and he soon joined the faculty as an assistant professor of biological statistics. He was promoted to associate professor in 1965, and became a professor of biological statistics in 1970. Searle has also been a visiting professor at Texas A&M University, Florida State University, Universität Augsburg and the University of Auckland. He has published several statistics textbooks and has authored more than 165 papers. Searle is a Fellow of the American Statistical Association, the Royal Statistical Society, and he is an elected member of the International Statistical Institute. He also has received the prestigious Alexander von Humboldt U.S. Senior Scientist Award, is an Honorary Fellow of the Royal Society of New Zealand and was recently awarded the D.Sc. *Honoris Causa* by his alma mater, Victoria University of Wellington, New Zealand.

The following interview, with Martin Wells of Cornell University, took place over a number of visits to the home of Professor Searle in Ithaca, NY in the Fall of 2007.

## 1. THE EARLY YEARS

**Wells:** Shayle, tell me a little about your early education.

**Searle:** As a small boy I was, so my mother often told me, in love with numbers and arithmetic. Apparently even before starting school I used to scrib-ble such things as $1 + 2 = 3$ in a book of wallpaper samples used as a scratch pad. And throughout most of my school days I was occasionally moved up a class because of being good at mathematics. (Classes were not governed by age, as in the U.S.A., but by ability.) Mind you, mathematics was not particularly rigorous or conceptual at the kindergarten-type school where I was for a year, nor during my two years at a grade school. In 1937 I started at a boarding school (for 8–14-year-old boys) where the teaching was very good, including mathematics. After five years I transferred to a high school where the teaching was generally bad, except for mathematics.

*Martin T. Wells is Professor, Departments of Biological Statistics and Computational Biology, Social Statistics and Statistical Science, Cornell University, Ithaca, New York 14853, USA (e-mail: mtw1@cornell.edu).*

**Wells:** Tell me about your undergraduate days.

**Searle:** It was in March 1945 when I started at University. I was to be at Victoria University College in Wellington (N.Z.'s capital) 120 miles south of my home town Wanganui. It was a college of the University of New Zealand, at that time, formally New Zealand's only university with students only at its four colleges and two agricultural colleges dotted around the country—half of them in each island.

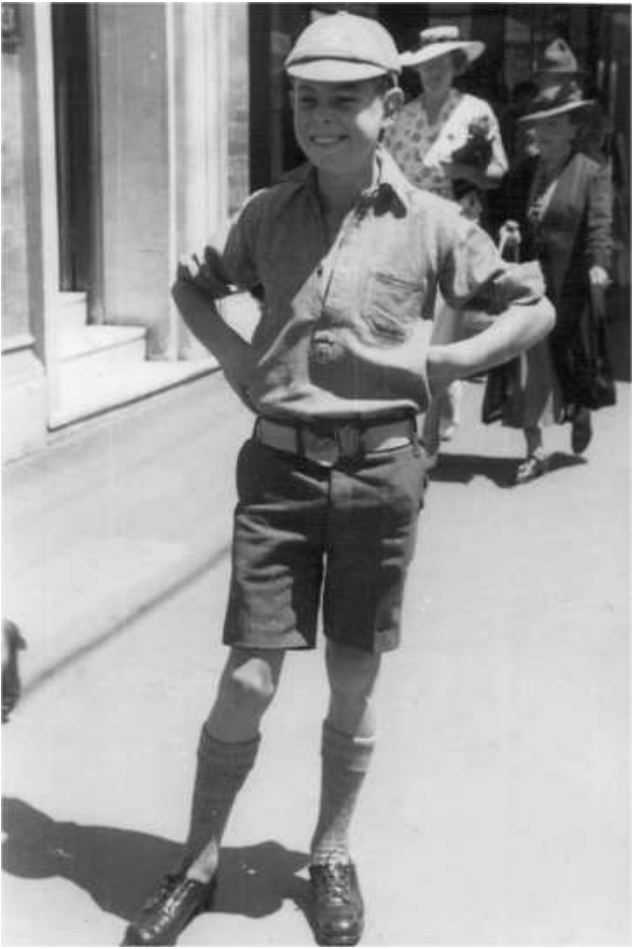

Fig. 1. *Shayle Searle, around 1937, school uniform, aged 9.*

They are now, and have been for some years, all autonomous universities, and the University of New Zealand has disappeared.

The difficulty I faced in 1945 was deciding what course of study I would follow. My family (no siblings, but a bunch of cousins) knew nothing about universities; least of all did they know anything about careers based on mathematics. Schoolmaster or accountant seemed the only options and because in my third and last year in high school (having the previous year passed the nationwide university entrance exam) I took and passed two year-long courses for the B.Com. degree, so I chose accountancy and spent half of that first university year as an office boy in a large accountancy firm in Wellington, and thus was a part-time student. The firm did a lot of auditing work and this led to my first apprenticeship, so to speak, of being an auditor: checking the arithmetic of long columns of journal entries in the books of a Lever Bros. soap-making plant near Wellington.

I found it to be incredibly dull work. That decided me; I wanted to do mathematics. So at year's end I quit my job, changed courses to do a B.A., and went home where I could, and did, get some excellent tutoring for three months to bone up on the maths I should have done in my third year at high school. All the work was algebra from an old and wonderfully good book by Hall and Knight.

**Wells:** How did you resolve these early career issues?

**Searle:** During that first year of mathematics I still had the crunch question: for what job would a mathematics training prepare me? Becoming an actuary came as the answer, surprisingly from a lady who owned a successful department store. With there being only four actuaries in New Zealand in the 1940s, none of whom were anywhere near my hometown of 20,000 people, basically a farming town, it was surprising any resident had even heard of an actuary! "Sort of a high-level accountant" was about the nearest description. Anyway, I found out about it, had another summer of tutoring and in May 1947 sat and passed the preliminary exam of the London Institute of Actuaries. The exam consisted of three hour papers in English and mathematics.

Then for the next two years I concentrated on the B.A. exams coming in 1948, these being two papers in pure maths and two in applied, the latter involving topics like statics, dynamics and hydrostatics: dull, difficult and for me from an agriculturally oriented background, of no use whatsoever. In 1949, after weathering a bout of pneumonia, I took the six exam papers for the M.A. in mathematics (no thesis required), one of which was on matrices. The instructor for that course was senior lecturer J. M. Campbell, using Aitken's 1948 book "*Determinants and Matrices.*" Campbell, a New Zealander as was Aitken also, had done his Ph.D. in statistics at Edinburgh where Aitken was having a very eminent career.

**Wells:** At that point, what path did you pursue after an undergraduate and masters training in mathematics?

**Searle:** After the 1949 M.A. exams I took a job as assistant to the actuary at Colonial Mutual Life Assurance Company in Wellington. I had no office of my own, but merely a desk in a large room with some dozen or so retirees who, day in and day out, were checking the weekly premiums paid for what were called industrial policies—something like twenty-five cents a week. The actuary's office was but a few steps across the hall. He was a real proper Englishman and helped me a great deal in preparing my first paper, "Probability: Difficulties of Definition." It was published in the *Journal of the Institute of Actuaries Students' Society*, 1951, pp. 204–212. Although I had read the von Mises book, and Venn's, I soon realized after attending my first lecture or two at Cambridge that my knowledge of probability was very naïve and incomplete. (I had not even had a course in set theory!)

Anyway, in 1950, now in the actuarial environment, I reverted to the actuarial exams, but still kept an eye on the B.Com. degree to which several courses in my B.A. degree (e.g., English and economics) could also be counted. So I took a statistics course for the B.Com. which also helped in preparing for Part I of the actuarial exams destined for May 1951. These and the following parts were known to be difficult; the average time for becoming fully qualified was eleven years! Individual exam questions dealt with annuities and life insurance premiums (with absolutely horrible notation) and some statistics. Many questions had such long descriptions that a paper took a full ten minutes to read—and after reading it one had to decide which questions to answer to satisfy the instruction "Do three from the five questions in each section of the exam paper"!! As if all this wasn't going to be difficult enough, there were no lectures available and only two or three books, some of them only in galley proof form. Notational distinction, in these books, of a population statistic from an estimator of it was sparse: often the same symbol was used for both. From England (where the exams originated) I was given the name and address of an actuary in an insurance company in Sydney, Australia who was supposed to be available to me to answer questions and give advice on my attempted solutions to exercises in the books. But, despite the almost daily flying-boat services between Wellington and Sydney, it usually took him a month to get his comments to me. Not much use. Anyway, in May 1951 I sat the exam.

**Wells:** How did you initially get interested in the subject of statistics?

**Searle:** In the 1949 M.A. exam I'd done much better than expected. By one mark out of 600 I was top of New Zealand; however, no kudos in that since there were only four examinees! Nevertheless, as a result, I became interested in an overseas scholarship to enable me to study statistics. Interest in statistics had been promoted by the course in Wellington and by the Part I actuarial exam. Unfortunately I discovered that I should have applied for the scholarship before, not after, my M.A. exams. I could apply after, but I knew I'd be competing with the notable New Zealander Peter Whittle (who has recently retired from his Cambridge professorship). So I scrubbed that idea. However, I was told that I could be supported overseas by a family agriculture trust (established by my successful maternal grandfather), so I proceeded to get myself accepted at both Emmanuel College and the statistics laboratory at Cambridge University.

## 2. CAMBRIDGE DAYS

**Wells:** Let's chat about your time at the statistics laboratory at Cambridge University. Tell me about your introduction to Cambridge University.

**Searle:** Departure from New Zealand in mid August, by ship, was not easy; three hours before leaving the Wellington wharves I received a phone call from the government actuary (the Institute's official representative) telling me that I had failed the whole of the Part I exam taken in May. That was a bitter pill to add to the emotion of a ship pulling out from its berthage for what was to be its usual 31-day voyage to Britain. My first days after arrival in Britain were spent in London during which time I went to see the Institute of Actuaries. Compared to the facilities I'd struggled with in New Zealand for trying to pass their exams, the Institute looked wonderful: a variety of lecture courses, some 80–100 students, and very nearby was a big Prudential Assurance building where a large number of actuarial students were employed. If becoming an actuary had still been my intention, I'd have been very envious. Coming to the Institute was the obvious thing to do.

But I was going to Cambridge. And a day or two after getting there, an easy hour by train, I paid a visit to the statistics laboratory. After my knock on the director's door, I followed "come in" and announced myself "Shayle Searle, from New Zealand." "Who are you?" said John Wishart (of Wishart's distribution). "I've never heard of you!" That did not seem to be a very auspicious start. However, in gentlemanly English manner, Dr. Wishart, said "Well, you've come a long way so we can't send you back."

**Wells:** What happened after this auspicious introduction at Cambridge?

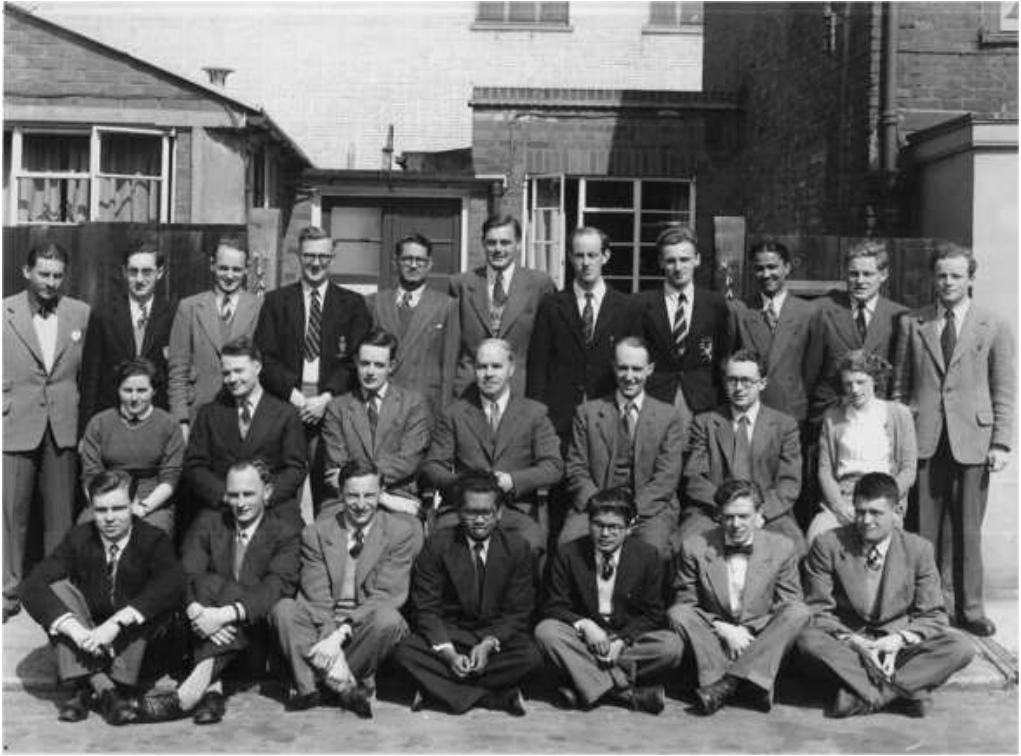

FIG. 2. *Cambridge University Statistical Laboratory 1953 personnel. Back row: D. A. East, B. Guss, E. S. Page, G. A. Coutie, K. K. Chaudary, D. J. Newell, F. J. Chatterly, J. R. Bell, R. S. Bawa, J. R. Ashford, P. A. Wallington. Second row: B. Reifenberg, W. L. Smith, D. R. Cox\*, J. Wishart (Director)\*, F. J. Anscombe\*, D. V. Lindley\*, P. A. Johnson. Front row: J. N. Darroch, B. D. Gee, W. S. Townson, K. W. C. de Silva, B. Das, G. B. Aneuryn-Evans, S. R. Searle. Absent: H. E. Daniels\*, Th. Metakides, J. T. Laws. (\* Faculty)*

**Searle:** I stayed; and from among the star-studded faculty of F. Anscombe, D. R. Cox, D. Lindley and H. E. Daniels I was given Dennis Lindley as my tutor and into whose class on probability I was directed. Boy, was I lost. But fortunately courses did not have exams; and my learning revolved around the customary 2–3 hour tutorial session I had each week with Lindley. Just he and me. Most of the time centered on my attempts at answering questions that came from previous years' exams for the Diploma. To begin with I was expecting to work for a Ph.D. But after a couple of months or so, Lindley told me like it was: he recommended that I do the Diploma and not the Ph.D. His reasoning was as follows: statistics has a formal connection to the math department and mathematicians do not always look very favorably at statistics. Yet they usually come to the final oral exam for statistics Ph.D. candidates. And often they decline to award the Ph.D. but instead award an M.A.—which in this situation has come to mean "Failed Ph.D." and there was no recourse. Lindley felt that this is what would happen to me, and as he rightly said, "you don't want to work for just a Failed Ph.D." Agreement was clear.

**Wells:** How did you handle this early disappointment?

**Searle:** During those early years I did have a disappointment or two: mis-timing an application for the overseas scholarship; getting no help preparing for the actuarial exams; the "Who are you?" introduction to statistics at Cambridge; and then being discouraged from the Ph.D. degree. You ask "how did I handle" all this? Certainly in those days the legions of counselors available today for all manner of situations did not exist. One largely relied on oneself and learnt to tough it out.

**Wells:** Tell me about your Cambridge Diploma project.

**Searle:** The Diploma has stood me well. It consisted of two papers, one theory and one data analysis, and also a written report resulting from being seconded for the academic year to a data-generating research project within the university: the report to describe the data analysis and its consequences. I was seconded to E. H. Callow's lab where he was measuring iodine number in the fat taken from different joints of various beef carcasses—and my efforts finished up as a co-author (see Callow and Searle, 1956). All this was usually considered a one-year effort, but in my case it was set at two years. At the end of the first year I sat the two exams for practice! My second year exam results were not as good as the first year; in fact of seven out of nineteen students who got "Passed with Distinction" yours truly was not among them. Lindley gave me the raspberry like I've never had it before or since. I'd been enjoying too much the social activities of college end-of-year festivities!

## 3. THE COMING OF AGE AS A STATISTICIAN

**Wells:** After Cambridge, what was your next move?

**Searle:** I then needed to find a job. Cambridge University had what I believe at the time (1953) was the early years of its career center. Although their advisors had clearly never previously dealt with a research (graduate) student, let alone one with statistics qualifications (and from New Zealand!), they did find me two interviews in London; one was for a job with the Colonial Service, in agriculture in Kenya (I resisted the temptation to ask the interviewer if his missing leg (or was it arm?) had been eaten by the Kikuyu), and the other was with Royal Dutch Shell who wanted to employ me in Venezuela. I decided to pursue neither opportunity when I heard of the possibility of a position in New Zealand, as a statistician at Ruakura Research Station, a large and comprehensive agricultural research farm. So I applied—but the position was canceled.

**Wells:** So much for the Cambridge career center; what did you do after the Ruakura Research Station job was canceled?

**Searle:** I returned to New Zealand and in October 1953 got a newly established post as Research Statistician with the Herd Improvement Department of the New Zealand Dairy Board, in Wellington. It turned out to be a decisive moment for my life's activities.

**Wells:** Tell me about your time at the New Zealand Dairy Board.

**Searle:** The work consisted of deriving ways of using dairy cow milk production records for deciding which cows and bulls would be used for breeding offspring (by artificial insemination) that would increase milk production not only for the individual farmer but for the nation also, since New Zealand has, for more than a century, lived by its exports of agricultural products; butter, cheese and milk powders being important parts thereof. The outstanding researcher in this discipline of animal breeding was Professor C. R. Henderson of Cornell University. And it was my good fortune that he came to New Zealand for his first sabbatical, and actually had a desk in my office for eight months from September 1955. His own Ph.D. from Iowa State University was in animal breeding, under the eye of Professor Lush, the father figure of the discipline. But Henderson had strong interests and training in statistics, and more than a nodding acquaintance with matrices. So we got on well together, especially after I showed him the formula for the inverse of a partitioned matrix needed in estimating environmental and genetic trends (see Henderson et al., 1959).

**Wells:** What was the consequence of your relationship with Professor Henderson?

**Searle:** The result of all this was that in August 1956 I went to Cornell and did a Ph.D. with Dr. Henderson. Before leaving New Zealand (with a Fulbright travel grant) I knew what my thesis topic would be, and by August 1958 had finished my Ph.D. That coincided with the New Zealand Dairy Board sending me data they wanted analyzed to investigate the possibility of having yearly production records estimated from just 3 or 4 months measured (sampled) production instead of the then-usual 9 months. Dr. Henderson was interested in this, too, and kindly kept me on as a Research Associate.

**Wells:** What did Cornell uniquely offer you as a graduate student?

**Searle:** I cannot describe Cornell's offerings as being unique because I have no comparison with other places since I applied nowhere else. But Cornell's tolerance of my special circumstances was wonderful: I arrived late, some two weeks into the semester, as a result of the travel arrangements made by Fulbright. Forming my degree committee was greatly aided by Henderson. Animal breeding was to be my major (with Henderson with his strongly statistical interests); one minor was to be statistics with Federer, head of Biometrics. The second minor was troublesome because I refused to do mathematics (I felt I had enough), and I couldn't do anything related to embryology because I had absolutely no background in chemistry or biology or physiology. Henderson came to the rescue by reassuring the department head to take me on with a minor of Animal Science doing a few undergraduate courses, in at least two of which (dairying and sheep husbandry) I gave the lectures on breeding. So I scrambled through!

Above all, the greatest benefit of Cornell was the complete freedom and encouragement to get on with what I wanted to do. I knew what my thesis topic was to be, I was getting the data for it from the N.Z. Dairy Board, the department had just got its own computer (an IBM 650), I wrote my own programs and worked many nights on the computer from 10 pm till 6 am. The freedom was superb—and productive.

**Wells:** After writing your Ph.D. with Professor Henderson what did you do?

**Searle:** I finished the Ph.D. at the end of 1958 and was hired as a Research Associate under Henderson, attending to an extension or two of my thesis, writing several papers for publication, and learning as much as I could about computing facilities needed for this kind of work. Henderson and I gave a semester-long seminar on unbalanced data and I wrote it up as an extensive set of notes, the proofreading of which was left to me.

**Wells:** What was your next move?

**Searle:** In late 1959 I returned to N.Z. and my position with the New Zealand Dairy Board where a sire-proving scheme was being inaugurated for selecting bulls to be sires in the artificial breeding program. For me it was a period of successful publication, for example, nine publications in 1961, not only in *The Journal of Dairy Sciences*, but in *Biometrics*, *Journal of Agricultural Science*, *Annals of Mathematical Statistics*. During this time I became a one-third-time scientist of the N.Z. Department of Scientific and Industrial Research in their Mathematics laboratory where I took part in their introducing computing and programming to the country's scientists.

1961 was also the year I was asked to reduce my research and spend time visiting dairy farmer meetings and giving talks. The happy coincidence was that, without my knowing it, I was being considered for a job at Cornell as statistician to their Computing Center. The official offer to me was delayed several months because two members of the committee deciding to employ me each thought the other had written to me. When the offer did come I of course accepted it to start on June 1, 1962 after finishing some responsibilities in New Zealand.

**Wells:** How did working at the N.Z. Dairy Board influence you?

**Searle:** Dealing with dairy cow production records made me realize that unbalancedness of data can materially affect the meaning of *many* of the calculations (e.g., sums of squares) that were being used in (at least agricultural) research literature. And this was before the flood of computer software that we have today. The Dairy Board work simply started me down the path of unbalanced data, matrices and variance components. Genetic studies use a ratio of variance components and that prompted estimating those components and that was highlighted by the 1953 Henderson paper in *Biometrics*.

**Wells:** Tell me about Henderson's influence on your work.

**Searle:** His greatest influence on me was his enthusiastic encouragement. For example, it was a custom in the Animal Science Department that each semester every graduate student had to be part of a team to give a seminar. In my first doing this I'd been allocated to talking about the uptake of iodine in the thyroid—something I knew absolutely *nothing* about. So for my remaining five semesters I suggested a topic on breeding to Henderson, he roped in another graduate student and the job got done. He was also very tolerant of my asking questions, and was exceedingly patient of my saying "I still don't follow you," and he would try again to placate me. He was also greatly helpful in suggesting improvements to whatever I was writing—although when it came to proofreading a supposedly final draft of a paper, a modicum of procrastination and delay would sometimes set in!

**Wells:** What was the state of random effects modeling in the 1950s?

**Searle:** It was quite limited: mostly for balanced data. And almost the only method of estimating variance components was what we today call the analysis of variance method. In its general terms it consists of equating sums of squares (or other quadratic forms) of data to their expected values, in which the random effects give rise to their variances. The trouble was that no real criteria were used for deciding on which sums of squares to use. With balanced data, analysis of variance seemed an "obvious" choice, and usually yielded as many sums of squares as variances being sought. But for unbalanced data there could be an excess number of sums of squares which made a problem for the desired estimation.

## 4. BACK TO CORNELL

**Wells:** When did you return to Cornell?

**Searle:** At Christmas time 1961 I received a letter from a friend at Cornell saying he was glad to hear that I was to be returning to Cornell. That was news to me; I'd heard nothing. Around March 1962 I wrote to Henderson to find out what the story was. It turned out that two members of the university computing committee each thought the other had written to me, but in fact neither had. So then I did get a letter; could I start in six weeks? I pointed out I was nine thousand miles from Cornell and my wife was expecting our second child, and I was already committed to some Dairy Board responsibilities, but yes, I could arrange to start on June 1st of that year, 1962.

**Wells:** What was your new position at Cornell?

**Searle:** In 1962 when I started in Cornell's Computing Center there was no commercial software available. Will Dixon and colleagues at UCLA were well on the way with BMDP package; but SAS had barely started (its first annual user conference was 1976) so part of my responsibility was to decide what statistical packages we should have and to get them programmed. The Computing Center had a statistical programmer who could do a credible job, and we proceeded to provide for regression and for analysis of variance of data from well-designed and executed experiments (i.e., balanced data).

**Wells:** What was the state of "modern computing" in 1962?

**Searle:** Computing in 1962 was rudimentary compared to today's activities. Cornell had begun in 1956 with an IBM 650 (2000 words of 10-digits plus sign) and in 1962 had a 1604 CDC. There was no commercial software, no data editing and few programming languages: Fortran and Algol. The consulting work was often quite elementary, such as correcting the following misadventures: regression analysis that used as data the $-1$s that had been entered in place of missing observations; the reproduction of data so that there were 800 of them because the 400 actual data were too few in number to make a correlation estimate be significant; the scrutinizing of some six pages of data for which a published analysis of variance seemed spurious; it was, because amongst 300 3-digit data we found two values had been entered as 5 digits (only 100 times too big!).

**Wells:** How did you get affiliated with the Biometrics Unit?

**Searle:** My consulting job also came with a courtesy appointment as assistant professor in the Biometrics Unit of Cornell's College of Agriculture as it was then named, but with ..."and Life Sciences" added to it later. This was where I had formally done the statistics part of my Ph.D. under the very helpful eye of Professor W. T. Federer, head of the Biometrics Unit. And that helpfulness and encouragement re-asserted itself on my joining the Biometrics Unit as faculty in 1962. I was enthusiastically urged to write up whatever I was working on. And I certainly did; five papers both that year and the next.

In 1965, just as computing was becoming a big item on campus, I accepted a line item assistant professorship in Biometrics and gave up my responsibilities as consultant at the Computing Center. The College of Agriculture started to have its own computing facilities and I became lightly involved with some aspects of that operation. But I had decided I wanted to be a statistician and not a computer-nic. That started my thirty years in the Biometrics Unit which revolved around three interrelated topics: matrix algebra, linear models and variance components estimation. For each of these three I started a course and wrote a book or two. Writing, to me, was an enjoyable form of hard work so I kept at it.

## 5. VARIANCE COMPONENTS, LINEAR MODELS AND MATRICES

**Wells:** What researchers showed an early interest in variance components?

**Searle:** In the 1950s only a small coterie of statisticians (many of them with animal breeding interests) felt comfortable with random effects. Occasional papers by such people as Crump, Daniels, Eisenhart, Winsor and Clark, Tippett, and Cochran made interesting but not earth-shattering contributions and mostly dealt with analysis of variance methods for balanced data. I remember, as a graduate student, being at a 6-week research gathering in 1957 called a seminar on analysis of variance held in Boulder, Colorado under the direction of Oscar Kempthorne with such notables as David (now Sir David) Cox, Bill Kruskal and Jerry Cornfield and others in attendance. Following my lecture there on variance components I had several people come up to me and ask me to "really explain random effects," one such being Jerry Cornfield. Well, after all, I suppose 1957 is half a century ago!

**Wells:** What computational issues were there in variance component modeling those days?

**Searle:** Not only were random effects not widely understood, but the computations were horrendous for unbalanced data. There was a series of papers giving scalar formulae for sampling variances of variance components estimates obtained from the analysis of variance method of estimation and on unbalanced data, but these formulae were incredibly complicated. And there was no software; indeed, in 1955 before going to Cornell, I struggled with a very small data set to do these calculations with a Powers-Samas punched card tabulator using the method of successive digiting (see Searle, 1993) for obtaining sums of squares and products. It was horrible.

**Wells:** How did you get interested in unbalanced data?

**Searle:** My strong interest in unbalanced data (having unequal numbers of observations in the subclasses of the data) arose from dealing with dairy production records when working for the Dairy Board. Herds do not all have the same number of cows, not all cows give milk every year, and within a herd varying numbers are of the same age. I clearly remember being puzzled for a long time in statistical methods giving two different least squares estimates of fixed effects in a one-way classification depending upon whether one assumed that one effect was zero, or that all the effects summed to zero. Even as late as my second year as a graduate student (1957) when Henderson and I gave a weekly 2–3-hour seminar on unbalanced data we were still confused by this situation.

**Wells:** How did the notion of the $g$-inverse change your thoughts on linear models?

**Searle:** One of our troubles was we had not kept up with the concept of estimability propounded by R. C. Bose in North Carolina [linear combinations of the parameters $\beta$, say $A\beta$, are defined as estimable if the rows of the matrix $A$ belong to the vector space spanned by the rows of the design matrix; Bose, 1949]. Nor were we aware of Penrose's (1955) generalized inverse matrix which, as Rao (1962) demonstrated, clarified the whole business of solving least squares equations which are so often not of full rank, and thus have an infinite number of solutions, but which, with the aid of a generalized inverse, easily lead, for every solution, to unbiased estimators of estimable functions. Some details of this situation are in my 1966 book *Matrix Algebra for the Biological Sciences*; they are considered more fully in *Linear Models* (1971).

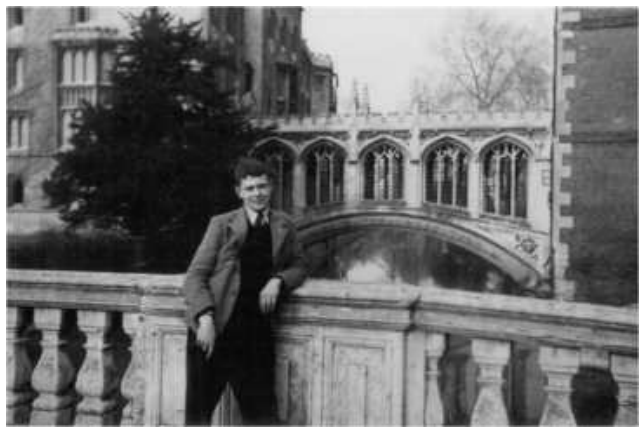

Fig. 3. *Shayle, 1952, on St. John's Bridge, Cambridge.*

**Wells:** You were an early advocate of using matrices in statistics; looking back this perspective seems obvious. Do you have a conjecture why early progress on the application of matrices to statistics was so slow?

**Searle:** The first of my *Annals* papers of 1956, 1958 and 1961 was "Matrix Methods in Variance and Covariance Components Analysis." Its title begs the question: Why has it taken so long for matrices to get widely adopted where they are so extremely useful? After all, matrices are two hundred and some years old and their use in statistics is only slowly becoming commonplace. But this was not so, even as recently as the 1950s. Even at Cambridge, in lectures on regression in 1952 there was no use of matrices. In Aitken's two 1939 books, one on matrices and one on statistics, neither mentions the main topic of the other! The very first paper in the first issue of *Annals of Mathematical Statistics* (Wicksell, 1930) is entitled "Remarks on Regression" yet it has no matrices. And even the Williams (1959) book on regression has only a tiny mention of matrices. Maybe this tardiness of adoption of matrices arose from their being treated so much a topic of pure mathematics that they remained hidden from their practicalities.

**Wells:** Tell me more about your early efforts in teaching linear models using matrices.

**Searle:** Around 1960 a visitor to the Biometrics Unit taught a course out of Graybill's excellent 1961 book *An Introduction to Linear Statistical Models*. He made very slow and pedantic progress and never got anywhere near the difficulties of unbalanced data. A year later D. S. Robson took on the course but after a few weeks had to be absent at research meetings and I was left with the teaching. In progressing toward the all-important result about a quadratic form in normal variables having a chi-squared distribution, Graybill (1961) had nineteen preparatory theorems! That struck me as just too much. To highlight the differences between each theorem and the next I summarized the nineteen in one line each. That immediately showed most of those differences to be very small; for example, normal variables with zero mean in one theorem had nonzero mean in the next. Among my biometry colleagues was a Ph.D. graduate of Graybill's who explained that was what Graybill wanted his students to learn and so be able in exams to regurgitate theorems and their proofs. Not for me, I decided. I wanted students to know where they could read the importantly useful theorems (which they might need to use in practice), and to thoroughly understand them. So I concentrated on the overriding theorem in this topic, namely the conditions under which a quadratic form of nonzero-mean normal variables has a noncentral chi-squared distribution. Armed with that, many of Graybill's nineteen theorems became just special cases. This appealed to me as a mathematically tidy way of handling things. Thus there was only one theorem, but a vital one, that students needed to know and in doing so needing to know that they understood it and knew how to use it. This set me to thinking about doing a book.

So then, armed with matrix algebra and the generalized inverse, and motivated by unbalanced data, I went on to describe in detail the various sums of squares and their corresponding hypotheses that can be derived from unbalanced data in the analysis of variance context. Not much of this was dealt with by Graybill or any other book. None of it was pretty, but it was only the use of matrices that made it at all feasible. As well as fixed effects models, *Linear Models* also (in its last three chapters) deals with applying to unbalanced data the analysis of variance method of estimating variance components, namely equating observed sums of squares to their expected values. Nearly all of that has now been relegated to history by the widespread application of maximum likelihood (starting with Hartley and Rao, 1967) and other methods, and the amazing growth of computability.

## 6. BOOK WRITING

**Wells:** You just mentioned that when teaching linear models that set you off to start thinking about doing a book. Tell me about writing your first book.

**Searle:** Federer was on sabbatic leave 1962–1963, and in his absence Professor D. S. Robson chaired the Biometrics Unit. In January 1963 he told me the secretaries were short of work, and he asked, "Don't you have some notes on matrices they could type?" I protested that although I'd written the notes for teaching a 1957 summer course when I was a graduate student, they needed plenty of work to make them worthy of a typist's time. Robson's reaction was, "Why don't you write a book?" So I did. I sent it in 1964 to four publishers: two turned it down, one never replied and Wiley & Sons accepted it. Months later they had a change of editors and turned it down. But luckily one of their senior editors, Ms. Beatrice Shube, saw the manuscript and promoted its publication. Thus was born my first book, Searle (1966) which sold more than 10,000 copies before going out of print. It spawned a mildly plagiarized version in the form of Searle and Hausman (1970) which through getting little or no promotion from business academia sold barely 5,000 copies. Nevertheless in 1974 it did have reprint editions in Taiwan (in English) and in Russia (in Russian), both of which were initially denied by their respective publisher. The successor to both the 1966 and 1970 books is *Matrix Algebra Useful for Statistics.* It (Searle, 1982) has sold more than 10,000 copies—thanks to George Styan for the "Useful." Prior to that helpful word, reviewers of the manuscript had strongly disliked the title.

I started, in 1965, and for thirty years taught a Matrix Algebra course at Cornell; it never had less than twenty students and up to seventy one year. They were from as many as 8–12 departments in agriculture most years, despite the 8:00 am starting time three days a week. Because of that early hour I never admonished anyone for being late; to be late was better than not coming. Thirty and more years ago teaching matrix algebra because of its practical use in statistics was never doubted as being useful (in contrast to some of the concepts of linear algebra). But nowadays, because, I suppose, of the computing software for doing the algebra, the teaching of the algebra seems to have become somewhat of just an add-on, if that. What a pity; matrix algebra is fun. My initial intrigue at being able to have $AB = 0$ without having $A = 0$ or $B = 0$ has never left me.

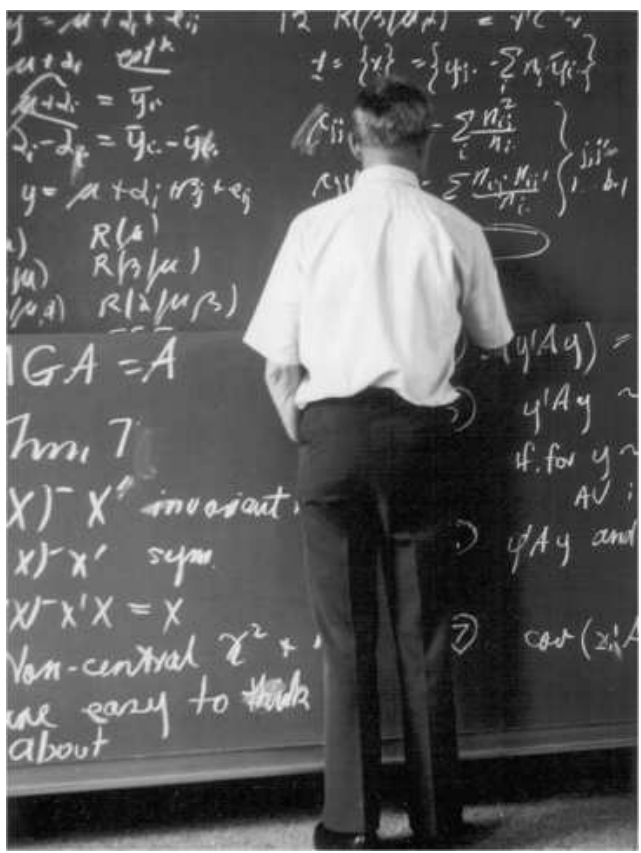

Fig. 4. *Shayle at the blackboard, Hoehenheim University, July, 1977.*

**Wells:** How did your classic *Linear Models* book come about?

**Searle:** In 1968 I was invited by H. O. Hartley to take my sabbatic at Texas A&M University and it was there that I started my Ph.D. level *Linear Models* book, published by Wiley & Sons in 1971. It has had sales of more than 15,000 and another 1,800 in the paperback Wiley Classic Edition which started in 1997. It is, I believe, a book which did make an impression on the understanding of linear models, especially of the complications emanating from unbalanced data. This was, and still is, especially important for using the high-powered computing software designed for doing linear model and analysis of variance calculations of such data—software such as *SAS, SPSS, STATA* and many others. Their early output descriptions and labels were often not a model of clarity, so that knowing the mathematics of the calculations was important.

The book led to many interesting and enjoyable invitations to give short courses for George Washington University in Washington, D.C. and in Berlin, Germany; and to lecture in such various locales as Budapest, Sydney, Auckland, and Freiburg am Breisgau and a raft of conferences and seminars in the U.S.A. and elsewhere. This included in each of 1985 and 1986 a 4-month stay in the mathematics department of the University of Augsburg in Bavaria, funded as a U.S. Senior Scientist by the Alexander von Humboldt Foundation of Bonn, Germany. As well as having a thoroughly enjoyable time in the historical city of Augsburg, I finalized a number of papers, gave some seminars and made a good start on *Linear Models for Unbalanced Data* published by Wiley in 1987.

To whatever extent all this represented success for *Linear Models* it motivated me to more books, five more actually, the most recent being *Generalized, Linear, and Mixed Models* co-authored with C. E. McCulloch, published by Wiley & Sons in 2001. An important feature of this book is its distinct emphasis on mixed models, a topic which is very much in evidence in today's statistical research. A contributing reason for this is that today's computing facilities can handle the arithmetic that is needed for coping with random effects when modeling unbalanced data.

## 7. SOME PERSPECTIVE

**Wells:** Looking back over your career, do you see a recurring theme?

**Searle:** I find it hard to believe that through my activities with animal breeding data it was more than fifty years ago when I was first trying to deal with random effects and variance components in unbalanced data. After all, half the genetic contribution to a cow's milk production comes from its sire but it is only a random half—and thus we have a random effect when including the effect of sire in a linear model for its daughter's milk production record. And this in turn gives rise to a variance component for the random effect. This has been a statistical interest of mine ever since the C. R. Henderson (1953) paper "Estimation of variance components..." in *Biometrics*. My contributions motivated by that paper are in the *Annals of Mathematical Statistics* in 1956, 1958 and 1961. My most recent effort on this topic is the 1992 Wiley book *Variance Components* with G. Casella and C. E. McCulloch.

**Wells:** Where do you see basic statistical research heading?

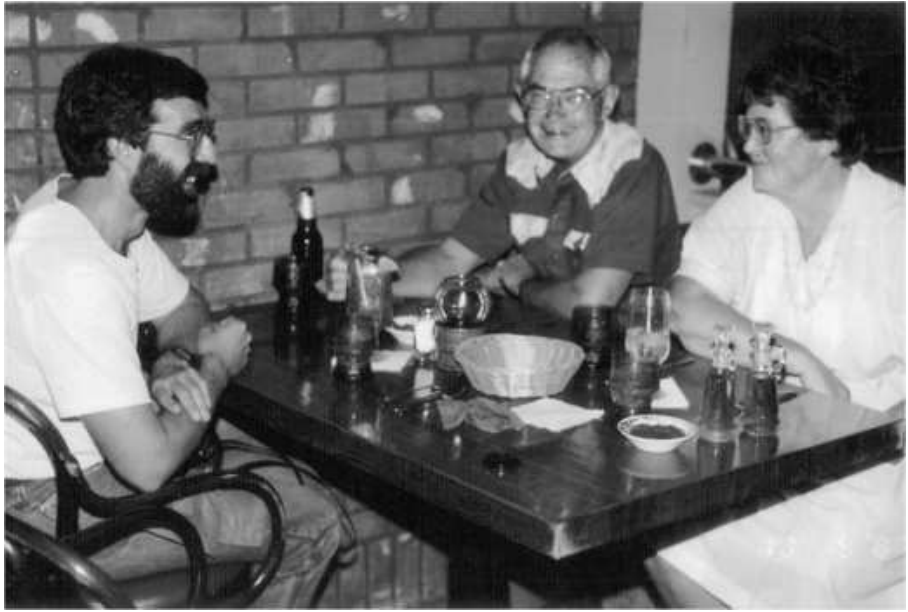

Fig. 5. *Shayle and his wife Helen enjoying the benefits of a conference with Harold Henderson in Bavaria, 1986.*

**Searle:** Certainly statistics research seems to have become increasingly computing oriented with great reliance (maybe indeed faith) being put on software. With this has also come diminishing interest in the algebraic development of new methods. Does this not arouse the questions "How will new methods be developed?" and "By whom?" And might not great reliance on software contain the possibility of very occasionally getting spurious output? These questions worry me. Especially so in the case of a student's own Ph.D. computer program that yielded several intelligible results from extensive data but also one completely outlandish result for which no adequate reason could be found. I insisted that it had to be a mistake in the student's own programming—but my insistence was eventually sidelined. That seemed to me to be not very good science.

## 8. RETIREMENT

**Wells:** I can speak for my Cornell statistics colleagues and let you know that we are sorry not to see you at the office more often these days.

**Searle:** Along with my own hip and knee replacement surgeries, my wife's illness and death, and the completion of two books, I slipped away from research, or perhaps more accurately research slipped away from me.

**Wells:** Tell me about some of your recent accolades.

**Searle:** Since retiring in 1995 I have had two very rewarding events bestowed upon me. In 1999 I was elected an Honorary Fellow of the Royal Society of New Zealand. The "Honorary" here indicates professional connection to New Zealand even when living and working beyond New Zealand. And the second event was the awarding in 2005 of the D.Sc. *Honoris Causa* by my alma mater, Victoria University of Wellington, New Zealand. Both events are acknowledged with gratitude.

**Wells:** As always Shayle, it has been delightful chatting with you. Thank you for granting me the opportunity to do this interview for *Statistical Science*.